\documentclass[letterpaper]{article} % DO NOT CHANGE THIS
\usepackage{aaai2026}  % DO NOT CHANGE THIS
\usepackage{times}  % DO NOT CHANGE THIS
\usepackage{helvet}  % DO NOT CHANGE THIS
\usepackage{courier}  % DO NOT CHANGE THIS
\usepackage[hyphens]{url}  % DO NOT CHANGE THIS
\usepackage{graphicx} % DO NOT CHANGE THIS
\usepackage{natbib}  % DO NOT CHANGE THIS AND DO NOT ADD ANY OPTIONS TO IT
\usepackage{caption} % DO NOT CHANGE THIS AND DO NOT ADD ANY OPTIONS TO IT
\usepackage{booktabs}  % in preamble
\usepackage{makecell}
\usepackage{algorithm}
\usepackage{algorithmic}
\usepackage{amssymb}
\usepackage{amsmath}
\usepackage{array}
\usepackage{newfloat}
\usepackage{listings}
\DeclareCaptionStyle{ruled}{labelfont=normalfont,labelsep=colon,strut=off} % DO NOT CHANGE THIS
\floatstyle{ruled}
\newfloat{listing}{tb}{lst}{}
\floatname{listing}{Listing}
\title{Generating Query-Relevant Document Summaries via Reinforcement Learning}

\author{
     Nitin Yadav\textsuperscript{\rm 1}, 
     Changsung Kang\textsuperscript{\rm 1}, 
     Hongwei Shang\textsuperscript{\rm 1}, 
     Ming Sun\textsuperscript{\rm 2}
 }
 \affiliations{
    \textsuperscript{\rm 1}Walmart Global Tech, Sunnyvale, California, USA\\
     \textsuperscript{\rm 2}Walmart Global Tech, Hoboken, New Jersey, USA\\
     nitin.yadav@walmart.com, changsung.kang@walmart.com, \\
     hongwei.shang@walmart.com, ming.sun0@walmart.com
 }

\usepackage{bibentry}
\begin{document}

\maketitle

%\begin{abstract}
%The title of a product, while crucial in e-commerce search, often fails to capture all the valuable information needed for search ranking. On the other hand, product descriptions are typically more comprehensive but too verbose and noisy to be directly used as input for ranking models, especially under strict latency constraints. This creates a critical need for a method to effectively condense product descriptions. In this paper, we propose a novel reinforcement learning framework that leverages search relevance as a reward to generate concise and optimized summaries of product descriptions. Experimental evaluations demonstrate the superior performance of our approach, showing significant improvements in both search relevance and engagement metrics.
%\end{abstract}

\begin{abstract}
E-commerce search engines often rely solely on product titles as input for ranking models with latency constraints. However, this approach can result in suboptimal relevance predictions, as product titles often lack sufficient detail to capture query intent. While product descriptions provide richer information, their verbosity and length make them unsuitable for real-time ranking, particularly for computationally expensive architectures like cross-encoder ranking models. To address this challenge, we propose ReLSum, a novel reinforcement learning framework designed to generate concise, query-relevant summaries of product descriptions optimized for search relevance. ReLSum leverages relevance scores as rewards to align the objectives of summarization and ranking, effectively overcoming limitations of prior methods, such as misaligned learning targets. The framework employs a trainable large language model (LLM) to produce summaries, which are then used as input for a cross-encoder ranking model. Experimental results demonstrate significant improvements in offline metrics, including recall and NDCG, as well as online user engagement metrics. ReLSum provides a scalable and efficient solution for enhancing search relevance in large-scale e-commerce systems.
\end{abstract}


\section{Introduction}
Modern day search engines typically consist of two stages: retrieval and re-ranking. For systems with huge amount of documents to search from, retrieval stage utilizes sparse retrieval methods such as BM25 \citep{robertson2009probabilistic} and dense retrieval based on bi-encoders \citep{magnani2022semantic,ni2022large,shan2023beyond,lin2024enhancing}. 
Since the re-rank stage processes orders of magnitude less documents compared to the retrieval stage, 
it can utilize more powerful models such as cross-encoders \citep{shang2025knowledge}, 
that unlike bi-encoders can encode query and document jointly via attention mechanism. 
Although cross encoders are state-of-the-art rankers %\citep{puthenputhussery2025large}
, 
they are limited by the number of input tokens due to the quadratic complexity of the attention mechanism in the transformer architecture and having to attend to both queries and documents together at runtime. Unlike bi-encoders, the feature computation cannot happen offline. Owing to that limitation, the product's search context for the cross-encoder models deployed in production is often limited to their titles alone \citep{zhao2025explainable} and sometimes a few attributes such as brand, color, and size \citep{Vo:Shan:2024:Know}. 

\begin{figure}[t]
\centering
\includegraphics[width=0.5\textwidth]{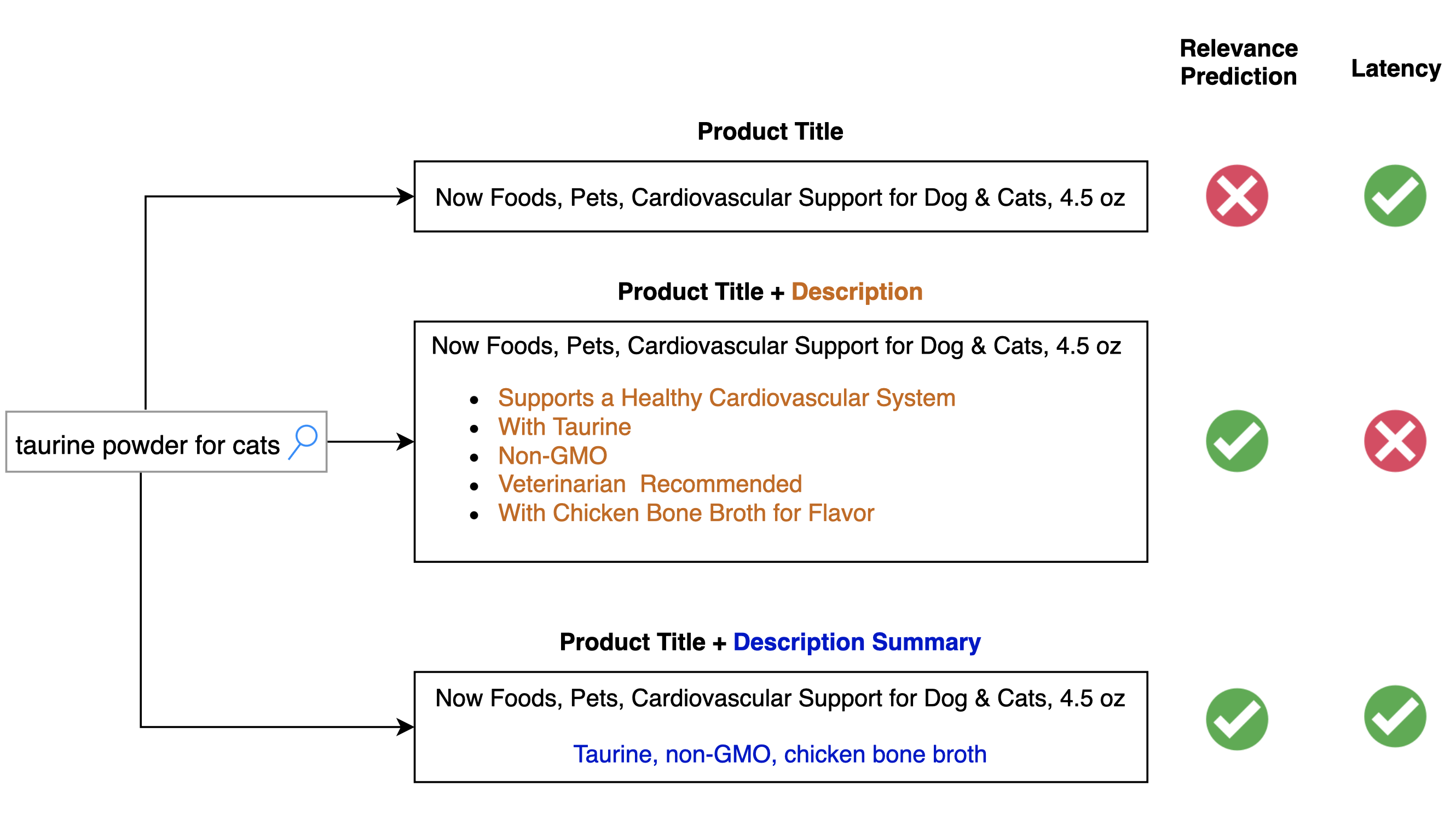} % Reduce the figure size so that it is slightly narrower than the column.
\caption{Tradeoffs between performance and latency depending on the size of product context showing the need to summarize the description}
\label{motivation}
\end{figure}

While using the product title alone as input for cross-encoder rankers ensures low latency, it often leads to suboptimal relevance predictions, as illustrated in the example query ``taurine powder for cats" as shown in Figure~\ref{motivation}. The product title, ``Now Foods, Pets, Cardiovascular Support for Dog \& Cats, 4.5 oz," lacks sufficient detail to match the query intent, resulting in poor relevance predictions. Adding the full product description, which includes details such as ``With Taurine," significantly improves relevance predictions. However, this approach is impractical in production due to the high latency caused by the quadratic complexity of the attention mechanism, especially since a typical full product description can be significantly longer than the one in this example.

Machine-generated summaries offer a practical solution to this problem. By summarizing the product description into concise, query-relevant attributes such as ``Taurine, non-GMO, chicken bone broth," we can retain the essential information needed for relevance prediction while minimizing the input token length. This enables cross-encoder rankers to achieve high relevance accuracy without incurring prohibitive latency costs, making them suitable for large-scale e-commerce search systems. Summarized descriptions strike a balance between relevance and efficiency, addressing the limitations of both sparse product titles and verbose full descriptions.

One way to summarize descriptions is to prompt LLMs to generate summaries, which may be reasonably good, but there is no guarantee that the generated summaries will be optimal for downstream tasks such as ranking, where customer queries are used as input alongside product information for the ranker.

Another way is to learn to produce queries as summaries, as in the Doc2Query framework \citep{nogueira2019document,nogueira2019doc2query,li2024doc2token}, but the setup of the problem is different compared to summarization. A set of queries is different from a summary, as there could be unnecessary repetitions of information between queries, with each query optimized/generated to match and retrieve the document. Additionally, this method shares the same issue as the simple LLM-based method mentioned above: the learning target (queries) is not completely aligned with the final downstream target (ranking).

A natural approach to solve this problem is to generate a single summary for each document or product and optimize the process using reinforcement learning, where the objective is to reward summaries that lead to improved performance on downstream tasks. In our case, the downstream metric of interest is search relevance. This paper explores and addresses this solution. Overall, our main contributions in this paper are as follows:
\begin{itemize}
\item Proposal of a reinforcement learning framework tailored for summarization with the goal of improving search relevance.
\item Strategy for constructing a training dataset facilitating efficient fine-tuning of summary generating models in that framework.
\item Comparison between Direct Preference Optimization (DPO) \citep{rafailov2023direct} and Group Relative Policy Optimization (GRPO) \citep{shao2024deepseekmath} for summarization.
\end{itemize}

The remainder of this paper is organized as follows.
Section \textbf{Related Work} provides an overview of existing research in the domain.
Section \textbf{ReLSum} formulates the problem and introduces the reinforcement learning framework for summarization.
Section \textbf{Experiments} presents details on the evaluation metrics and the results of offline and online testing.
Finally, Section \textbf{Conclusion and Future Work} summarizes the key contributions of the paper and outlines potential directions for future research.

\section{Related Work}
\label{sec:relatedwork}
One line of work to expand/enrich the document is Doc2Query \citep{nogueira2019document,nogueira2019doc2query}.
This method uses a sequence-to-sequence transformer to generate synthetic queries for each document, based on (query, relevant document) training pairs. These predicted queries are appended to the document text to enrich it, enabling improved retrieval performance.
To generalize Doc2Query in the e-Commerce search application, \citet{li2024doc2token} proposed Doc2Token, 
it predicts and appends relevant missing tokens--rather than full predicted queries--to a document, improving retrieval efficiency by avoiding redundancy in domains like e-commerce search. However, both Doc2Query and Doc2Token tend to generate more popular queries or tokens due to their learning target, which is not optimal for summary generation purposes. Ideally, we want summaries that can effectively support both popular and tail queries, ensuring broader coverage and improved relevance across diverse search scenarios.

Recently, the emergence of LLMs has pushed the field of text summarization into a new era \citep{zhang2025systematic,basyal2023text},
by introducing new modeling approaches with LLMs primarily in zero-shot 
and few-shot settings. These efforts explore a variety of techniques
to improve the quality of generated summaries,
including prompt-based \citep{ravaut2023promptsum,chang2023booookscore,sun2024prompt}, 
multi-agent \citep{zhang2023summit,ma2024iterative}, and alignment-based methods \citep{ouyang2022training,fetahu2023instructpts}. 

Although LLM-based summarization has received significant research attention,
using these techniques to augment documents with summaries does not necessarily
improve ranking performance in e-commerce search. This is because the objectives 
of summarization and ranking are not fully aligned. 
A solution for the alignment problem would be to jointly learn to summarize documents and assess query-document relevance. At a high level, a document's description could be passed to a trainable LLM 
to generate a summary, which could then be fed alongside the query, document title,
and other attributes into another trainable model that predicts query-document relevance. This setup has a challenge that direct backpropagation would not be possible due to the non-differentiable sampling steps in text generation. One workaround would be to use the Gumbel softmax trick \citep{jang2016categorical}, a differentiable approximation to sampling, as demonstrated by \citet{wichers2024gradient}. Another key limitation to this approach is that it requires the LLM and the downstream relevance model to share a common vocabulary.

With recent advancements in reinforcement learning (RL), researchers have begun applying RL to query augmentation, such as in DeepRetrieval \citep{jiang2025deepretrieval}—a method that trains large language models to generate queries by optimizing retrieval metrics (e.g., NDCG) as rewards. 
This approach outperforms prior methods across diverse tasks. 
However, to the best of our knowledge, there has yet to be work leveraging RL-based mechanisms
for document expansion in e-commerce applications. 

\begin{figure}[t]
\centering
\includegraphics[width=0.5\textwidth]{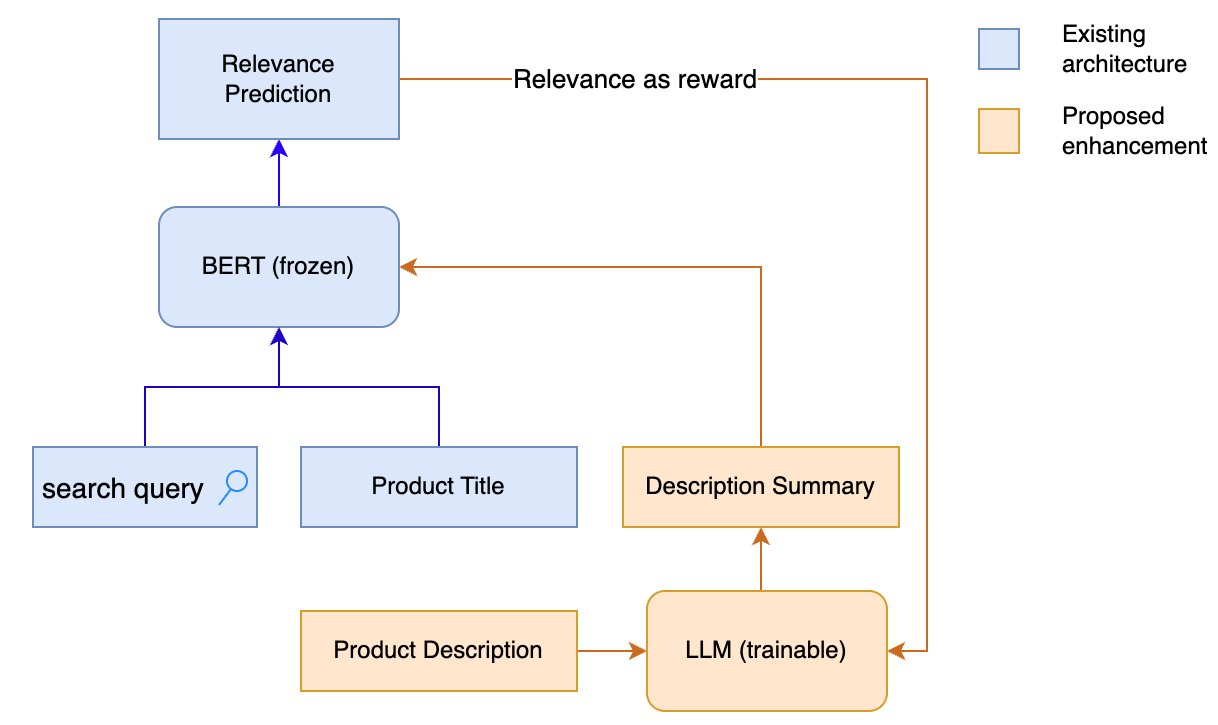} % Reduce the figure size so that it is slightly narrower than the column.
\caption{Proposed architecture. The blue components represent the current production system, while the orange components illustrate the enhancements introduced by our new approach. The trainable large language model (LLM) generates concise description summaries from product descriptions, which are optimized for relevance prediction. These summaries, along with the product title and search query, are fed into a frozen BERT-based cross-encoder model to evaluate alignment and predict relevance. Relevance scores are used as rewards to train the LLM, ensuring alignment between summarization and ranking objectives while addressing latency and vocabulary constraints.}
\label{arch}
\end{figure}

Our production ranking model 
% \citep{shang2025knowledge, puthenputhussery2025large} 
is a BERT \cite{devlin2019bert} model based on a cross-encoder architecture (shown as the blue components in Figure~\ref{arch}). 
We concatenate the query and the product title and feed the combined input into the BERT model. 
The hidden state corresponding to the [CLS] token represents the query-product pair. An MLP layer is then applied to this representation to produce a single relevance score.
It is important to note that the computational complexity of the BERT-based cross-encoder model grows quadratically with the number of input tokens. This is due to the self-attention mechanism, in which every token attends to every other token. 
As the model is computationally expensive, latency reduction techniques such as caching precomputed scores for popular query-product pairs and tokenizing the product title offline are employed to optimize performance. % \citep{puthenputhussery2025large}.
To meet latency constraints, the full product description is excluded from the model's input in the production environment.

\section{ReLSum: Summarizing Product Description for Search Relevance with Reinforcement Learning}
\label{sec:ReLSum}
\subsection{Proposed Architecture}
The proposed architecture, illustrated in Figure~\ref{arch}, addresses the challenge of connecting summarization directly with the downstream task of relevance ranking. Traditional approaches often rely on short product titles or long full descriptions, which either lack sufficient detail or incur high latency during ranking. This architecture introduces a novel solution by leveraging a trainable large language model (LLM) to generate concise description summaries optimized for relevance prediction.

These machine-generated summaries are designed to retain the essential attributes of the product while minimizing verbosity. The summaries are then used as input to a frozen BERT-based cross-encoder ranking model \citep{puthenputhussery2025large}, which evaluates their alignment with search queries and provides relevance scores as rewards. By incorporating relevance as a reward signal, the architecture ensures that the generated summaries are tailored to improve ranking performance.

Our approach addresses the limitations of prior methods discussed in the related work section, such as the misalignment between intermediate learning targets (e.g., queries or tokens) and the final ranking objective. Furthermore, compared to the Gumbel softmax trick \citep{jang2016categorical, wichers2024gradient}, our proposed architecture provides a more elegant and practical solution to align the output of an LLM with the downstream relevance model. By treating the relevance model as a frozen component and using its output as a reward signal for training the LLM, our approach avoids the constraints of requiring a shared vocabulary between the LLM and the relevance model, making it more suitable for real-world production systems. 
\subsection{Problem Formulation}
We formulate our problem as learning to summarize product description which when included in the search context of the product, maximizes the relevance of the product with respect to the search queries.

The current search system uses the user query $q$ and the product title $t$ as input for the relevance ranking model $r(q, t)$, which returns a relevance score between 0 and 1 with higher value indicating higher relevance. When a full product description $d$ or a product description summary $s$ is added to the item title, we represent it as $[t; d]$ and $[t; s]$, respectively.

For the example in Figure~\ref{motivation}, we would have:
\begin{figure}[h!]
\centering
\begin{tabular}{|p{0.9\linewidth}|}
\hline
\vspace{0.5mm}
$t$: Now Foods, Pets, Cardiovascular Support for Dog \& Cats, 4.5 oz \\
\vspace{0.5mm}\\\hline
\vspace{0.5mm}
$d$:
\begin{itemize}
    \item Supports a Healthy Cardiovascular System
    \item With Taurine
    \item Non-GMO
    \item Veterinarian Recommended
    \item With Chicken Bone Broth for Flavor
\end{itemize} \\ \hline
\vspace{0.5mm}
$s$: Taurine, non-GMO, chicken bone broth \\ \vspace{0.5mm}\\\hline
\end{tabular}
\label{example-box}
\end{figure}

Given a dataset $D\{(q_i,[t; d]_i,l_i)\}_{i=1}^N$ of queries, products and true relevance label $l \in [0, 1]$, our goal is to generate description summaries $s$ such that $r(q, [t; s])$ is close to $l$. Note that computing $r(q, [t; d])$ is prohibitively expensive at runtime in our setup.

\subsection{Reinforcement Learning (RL) Formulation}

We do not have supervision data of what these summaries $s$ should look like - making the problem ideal for reinforcement learning with the objective of minimizing the absolute difference between $r(q, [t; s])$ and $l$. Formally, we want to learn a policy $\pi_\theta$ parameterized by $\theta$ that generates $s$ given $[t; d]$ and maximizes

\begin{align}
\mathbb{E}_{\substack{s \sim \pi_{\theta}([t; d]) \\ (q, [t; d], l) \sim D}} \left[ -\left| r(q, [t; s]) - l \right| \right].
\end{align}

In this formulation, the policy model $\pi_\theta$ has access only to product information, while the customer queries and the search ranking system correspond to the \textit{environment} in reinforcement learning terminology. The environment observes the queries and provides relevance scores by the ranking model as rewards, enabling the policy to optimize its behavior. This formulation introduces a novel approach by constraining the policy model to rely solely on product information, a practical design choice that ensures scalability. By limiting the policy to product data, we can train the model on available query-product relevance data and generalize it to billions of products that lack query-specific labels. This approach allows for efficient deployment across large-scale e-commerce platforms while maintaining the ability to generate high-quality product summaries.

With $- \left| r(q, [t; s]) - l \right|$ as reward, any reinforcement learning framework can be used to find the optimal policy. We use GRPO \citep{shao2024deepseekmath} and update the policy $\pi_\theta$ that maximizes

\begin{align}
\mathbb{E}_{\substack{
  \{s_i\}_{i=1}^G \sim \pi_{\text{old}}([t; d]) \\
  (q, [t; d], l) \sim D
}} \left[
\frac{1}{G} \sum_{i=1}^{G} \frac{1}{|s_i|} \sum_{j=1}^{|s_i|} 
\left( \text{MCA}_{i,j} - \beta\, D_{\text{KL}}\left[ \pi_{\theta} \,\|\, \pi_{\text{ref}} \right] \right)
\right]
\label{eq:grpo}
\end{align}

\noindent , where $\pi_{old}$ is the policy used to sample the summaries $\{s_i\}_{i=1}^G$ at each training step, $\pi_{ref}$ is the initial policy which does not change throughout the training process, and 
$\text{MCA}_{i,j}$ (min-clipped advantage) is defined as

\[
\begin{split}
\text{MCA}_{i,j} = \min \Bigg[ & \dfrac{\pi_{\theta}(s_{i,j}|[t; d],s_{i,<j})}{\pi_{old}(s_{i,j}|[t; d],s_{i,<j})} A_i, \\
& \text{clip} \left( \dfrac{\pi_{\theta}(s_{i,j}|[t; d],s_{i,<j})}{\pi_{old}(s_{i,j}|[t; d],s_{i,<j})}, 1 - \epsilon, 1 + \epsilon \right) A_i \Bigg]
\end{split}
\]
, where
\[
A_i = \dfrac{(-\left| r(q, [t; s_i]) - l \right|) - mean(\{ (-\left| r(q, [t; s_i]) - l \right|)\}_{i=1}^G )}{std(\{ (-\left| r(q, [t; s_i]) - l \right|)\}_{i=1}^G)}.
\]
and $\epsilon$ is a clipping-related hyper-parameter for stabilizing training. 
In the remainder of the paper, we will refer to the model finetuned with the above objective as ReLSum$_{\text{GRPO}}$.

\subsection{DPO Formulation}
Another way to optimize the policy is through direct alignment algorithms such as DPO \citep{rafailov2023direct}. Using DPO, policy $\pi_\theta$ is updated to maximize
\begin{equation}
\begin{split}
\mathbb{E}_{ \{s_i, s_k\} \sim \pi_{\theta}([t; d]), (q, [t; d], l) \sim D} 
& \Bigg[ \log\sigma\Bigg( \beta\log\dfrac{\pi_{\theta}(s_i|[t; d])}{\pi_{ref}(s_i|[t; d])}\\ 
& - \beta\log\dfrac{\pi_{\theta}(s_k|[t; d])}{\pi_{ref}(s_k|[t; d])} \Bigg) \Bigg]
\end{split}
\label{eq:dpo}
\end{equation}

\noindent such that

\[
\left| r(q, [t; s_i]) - l \right| < \left| r(q, [t; s_k]) - l \right|.
\]
In the remainder of the paper, we will refer to the model finetuned with the above objective as ReLSum$_{\text{DPO}}$.

\subsection{Training Dataset}
To optimize the training signal for the RL and DPO formulations outlined in Equations \ref{eq:grpo} and \ref{eq:dpo}, we construct a training dataset in which the description provides critical information relevant to the query, while the product title alone does not suffice. The process for collecting this dataset is outlined as follows:
\begin{enumerate}
\item We begin by gathering a large dataset consisting of queries and corresponding products sourced from the search logs of an e-commerce platform.  
\item Since the dataset is too large to assign human-judged relevance labels ($l$) manually, we compute $r(q, [t; d])$ offline as a proxy for relevance labels ($l$).
\item To focus on training samples where the product description significantly influences the relevance, we filter the dataset to retain instances with a noticeable difference between the relevance score using the description, $r(q, [t; d])$, and the relevance score without the description, $r(q, t)$. This filtering ensures that the dataset emphasizes examples where a well-crafted summary, one that effectively addresses the query, receives a reward that distinguishes it from less optimal summaries.
\end{enumerate}

\subsection{Model Training}

We use an LLM as the model for generating the description summaries and an existing fine-tuned BERT model as the reward model, which is trained as a cross-encoder model. Model training phase involves training only the LLM; reward model is kept frozen throughout the process. As shown in the Figure \ref{arch}, each training step consists of:
\begin{enumerate}
    \item Prompting LLM to summarize the product description, specifically asking to add product attributes from description that are not present in the title. We found the following instruction to be quite effective in generating good summaries:
    
\begin{figure}[h!]
\centering
\begin{tabular}{|p{0.9\linewidth}|}
\hline
\vspace{0.5mm}
[DESCRIPTION]: $\langle description \rangle$  [TITLE]: $\langle title \rangle$ Product attributes appearing in [DESCRIPTION] but not in [TITLE] are: \\
\vspace{0.5mm}\\\hline
\end{tabular}
\label{inst-box}
\end{figure}

    Here $\langle description \rangle$ and $\langle title \rangle$ are the actual description and title of the product respectively.
    \item We sample $G$ responses (summaries) from LLM for each prompt which is two in case of DPO and can vary for GRPO (we experimented with 4 and 8 but found 4 to give a good balance of training signal and training time).
    \item These response summaries are appended to the product title and along with the query used as an input to the reward model (BERT-based cross-encoder model) to compute the rewards.
    \item Rewards thus computed are in turn used to compute gradients to update the LLM's parameters as per GRPO or DPO objectives detailed in Equations \ref{eq:grpo} and \ref{eq:dpo}.
\end{enumerate}

\section{Experiments}
\label{sec:experiments}
We use \textit{Mistral-7B-Instruct-v0.3} \cite{mistral7binstruct} as the LLM for our implementation and train its LoRA \cite{hu2022lora} parameters using the AdamW \cite{loshchilov2017decoupled} optimizer. We experimented with different hyper-parameters for training but found the ones shown in Table~\ref{tab:hyper} to be the most optimal in terms of training time and model performance. During evaluation, we append the original full description or the generated summary as applicable to the product's title in the test dataset. In essence, we are only evaluating the input (either existing or generated) to the cross-encoder model.

\begin{table}[ht]
\small
\setlength{\tabcolsep}{6pt}  % Adjust spacing as needed
\centering
\begin{tabular}{lcc}
\toprule
\textbf{Hyperparameter} & ReLSum$_{\text{GRPO}}$ & ReLSum$_{\text{DPO}}$ \\
\midrule
LoRA rank               & 32    & 32    \\
LoRA alpha              & 32    & 32    \\
$\beta$ (KL coefficient) & 0.0   & 0.1   \\
$\epsilon$ (Clipping parameter) & 0.2   & N/A   \\
$G$ (Generated samples) & 4     & 2     \\
Temperature             & 0.9   & 0.9   \\
Learning rate           & $10^{-5}$   & $10^{-5}$   \\
Learning rate scheduler & cosine & cosine \\
Epochs                  & 1     & 1     \\
Batch size              & 8     & 8     \\
\bottomrule
\end{tabular}
\caption{Hyperparameters for Model Training}
\label{tab:hyper}
\end{table}

\subsection{Test dataset}

\begin{table}[ht]
\small
\centering
\begin{tabular}{c l c}
\toprule
\makecell{\textbf{Rating} \\ \textbf{Score}} & \textbf{Description} & \makecell{\textbf{Relevance} \\ \textbf{Label}} \\
\midrule
4 & Excellent – perfect match     & 1   \\
3 & Good – one attribute mismatched    & 0.5 \\
2 & Okay                          & 0   \\
1 & Bad – irrelevant              & 0   \\
0 & Embarrassing                  & 0   \\
\bottomrule
\end{tabular}
\caption{Rating score to relevance label mapping}
\label{tab:rel_label}
\end{table}

Our test data, which we refer to as golden dataset, consists of queries, products and rating scores $\in \{0, 1, 2, 3, 4\}$, obtained from an e-commerce platform's search logs
% \textit{Walmart.com}'s search logs 
with rating scores provided by human judges as shown in Table~\ref{tab:rel_label}. The table also shows the relevance labels ($l$) corresponding to the rating scores, which are also the targets for reward model ($r$) training. The full golden dataset consists of 4,781 queries and 133k query-product pairs, representing a sample of search traffic from an e-commerce platform.
% Walmart.com. 
A subset of this dataset, which we refer to as the golden-tail dataset is a sample of the 3rd tertile of the search traffic when sorted from highest to lowest traffic. We are interested in the golden-tail dataset because it tends to contain queries that are not very common and sometimes need product description to be answered. The golden-tail dataset consists of 2450 queries and 47k query-product pairs.

\subsection{Offline Results}

\begin{table}[ht]
\small
\setlength{\tabcolsep}{4pt}  % reduce column spacing
\centering
\begin{tabular}{lcccc}
\toprule
 & \multicolumn{2}{c}{\textbf{Golden-full}} & \multicolumn{2}{c}{\textbf{Golden-tail}} \\
\cmidrule(lr){2-3} \cmidrule(lr){4-5}
\textbf{Method} & \textbf{R@90P} & \textbf{NDCG@5} & \textbf{R@90P} & \textbf{NDCG@5} \\
\midrule
None              & 0.00\% & 0.00\% & 0.00\% & 0.00\% \\
Desc              & 0.03\% & 0.33\% & 0.35\% & 0.50\% \\
$\text{Sum}_{\text{ref}}$   & -0.01\% & 0.39\% & 0.27\% & 0.59\% \\
ReLSum$_{\text{GRPO}}$ & 0.13\% & 0.80\% & 0.51\% & 1.03\% \\
ReLSum$_{\text{DPO}}$ & 0.17\% & 0.68\% & 0.45\% & 0.97\% \\
\bottomrule
\end{tabular}
\caption{Gains in percentage over the baseline (mentioned as None)}
\label{tab:offline}
\end{table}

For all our offline experiments, we evaluate the following candidates. Note that all the inputs are truncated beyond certain number of tokens irrespective of the candidates.
\begin{enumerate}
\item $\mathbf{None}$: No extra input other than the product title is added to the product context, which is our baseline.
\item $\mathbf{Desc}$: Description is added to the product context.
\item $\mathbf{Sum_{ref}}$: Summarized description by LLM without fine-tuning also known as $\pi_{ref}$ in equations \ref{eq:grpo} and \ref{eq:dpo} is added to the product context.
\item ReLSum$_{\text{GRPO}}$: Summarized description by LLM finetuned with GRPO is added to the product context.
\item ReLSum$_{\text{DPO}}$: Summarized description by LLM finetuned with DPO is added to the product context.
\end{enumerate}

\subsubsection{Metrics}
We use the following metrics for the offline tests on the golden datasets:

\begin{enumerate}
\item $\mathbf{R@90P}$: Recall at 90\% precision, which measures the recall of perfect-match products (products with a rating score of 4) while ensuring a precision threshold of 90\%. This metric is particularly important because, in the production system, the ranking score generated by the cross-encoder model is not only used for ranking but also serves as a classifier to filter out irrelevant products from the search results. For business reasons, maintaining high recall while ensuring reasonably high precision is critical to avoid excluding relevant products while keeping irrelevant ones off the search page. Thus, R@90P is one of the most important metrics for evaluating system performance.
\item $\mathbf{NDCG@5}$: Normalized Discounted Cumulative Gain at the top 5 positions, which evaluates the ranking quality by considering both the relevance and the order of the top-ranked products. This metric is crucial because customer engagement (e.g., clicks, add-to-cart actions, and orders) is most likely to occur within the top 5 positions on the search results page. Ensuring that the most relevant products appear in these positions is essential for maximizing user satisfaction and business outcomes.
\end{enumerate}

Results from the offline evaluations are presented in Table~\ref{tab:offline}. Both ReLSum variants outperform the rest of the candidates. Specifically, ReLSum$_{\text{GRPO}}$ achieves the best overall performance, with a +0.13\% gain in R@90P and a +0.80\% gain in NDCG@5 on the Golden-full dataset, and a +0.51\% gain in R@90P and a +1.03\% gain in NDCG@5 on the Golden-tail subset. ReLSum$_{\text{DPO}}$ also demonstrates strong performance, slightly trailing behind ReLSum$_{\text{GRPO}}$. Even though the magnitude of these gains may seem small, gains over 0.5\% in NDCG@5 in Golden-full dataset for example have resulted in significant improvement in user engagement metrics historically. 

In contrast, conventional summaries (Desc and Sum$_\text{ref}$) yield only marginal improvements, particularly on the Golden-tail dataset, and under perform compared to ReLSum. These results highlight the effectiveness of ReLSum in improving both recall and ranking quality. 

Notably, the gains are more pronounced in the tail segment of the test dataset. This aligns with our intuition that product descriptions are especially helpful for queries seeking specific attributes or details—types of queries that are more prevalent in the tail segment. By ensuring high recall at a precision threshold, ReLSum effectively addresses the dual objectives of relevance ranking and irrelevant product filtering.

\subsection{Online Results}
Encouraged by the excellent offline results, we initiated online experiments by deploying ReLSum$_{\text{GRPO}}$ in the production test environment. Our online experiments involved two tests:

\subsubsection{Interleaving}
Interleaving \cite{joachims2003evaluating} is an evaluation methodology 
% employed at Walmart 
to compare the number of add-to-cart (ATC) actions generated by rankings produced by a control model and a variation model. During the evaluation period, rankings from the control (production) and variation (ReLSum$_{\text{GRPO}}$) models were interleaved and presented to users. The performance metric used in this analysis was \textbf{ATC@40}, which quantifies the number of ATC actions within the top 40 ranked results for both models. The experimental findings revealed an ATC lift of \textbf{0.22\%} with a p-value of 0.000, indicating a statistically significant improvement in user engagement performance with the proposed model.

\subsubsection{Online A/B Test}

\begin{table}[ht]
\small
\setlength{\tabcolsep}{6pt}  % Adjust column padding
\centering
\begin{tabular}{lcc}
\toprule
\textbf{Metric}     & \textbf{Lift} & \textbf{P-value} \\
\midrule
GMV         & +0.29\%  & 0.1544  \\
Units       & +0.34\%  & 0.0112 \\
Orders      & +0.26\%  & 0.0193 \\
Converted Visits  & +0.14\%  & 0.0867  \\
\bottomrule
\end{tabular}
\caption{A/B test results with gains over controls in percentages}
\label{tab:ab}
\end{table}

We also performed an A/B test to measure the user engagement where users were split into two groups, with one group treated with the search results produced from the control (production) model and the other group treated with the search results produced from the variation (ReLSum$_{\text{GRPO}}$) model. 
This test was conducted over a 14 day period across all the segments in the search traffic. 
These two groups were then compared on the following metrics with the results shown in the table~\ref{tab:ab}.
The A/B test results in Table~\ref{tab:ab} show consistent improvements across key business metrics.
\begin{enumerate}
\item \textbf{GMV}: Gross-merchandise value per visitor
\item \textbf{Units}: Number of units in a completed order per visitor
\item \textbf{Orders}: Number of orders per visitor
\item \textbf{Converted Visits}: Number of visits with at least one completed order divided by total visits
\end{enumerate}

The results from both the interleaving and A/B experiments further validate the effectiveness of our proposed approach in improving search relevance and user engagement. Notably, products that were previously ranked low or filtered out by the classifier in the control setting—due to a lack of sufficient information in their titles to signal relevance—are now being ranked higher and retained in the result set. This is achieved because the generated summaries enrich these products with critical information that was previously missing, allowing the ranker and classifier to better assess their relevance to the query. The consistent gains in key metrics, such as ATC@40 and business performance indicators (GMV, Units, Orders, and Converted Visits), underscore the significant impact of our approach in delivering more relevant results and enhancing the overall shopping experience for users.

\begin{table*}[ht]
\small
\setlength{\tabcolsep}{6pt}
\centering
\begin{tabular}{>{\raggedright\arraybackslash}p{0.12\linewidth} >{\raggedright\arraybackslash}p{0.22\linewidth} >{\raggedright\arraybackslash}p{0.4\linewidth} >{\raggedright\arraybackslash}p{0.15\linewidth}}
\toprule
\textbf{Query} & \textbf{Title} & \textbf{Description} & \textbf{Summary} \\
\midrule
loreal revitalift \textbf{sunscreen} & Revitalift Anti-Wrinkle Firming Day Cream by LOreal Paris for Women - 1.7 oz Cream & Revitalift Anti-Wrinkle Firming Day Cream is a unique daily moisturizer with broad spectrum UVA/UVB SPF 18 \textbf{sunscreen} protection, Revitalift Complete Cream SPF 18 delivers anti-wrinkle and firming action for radiant, younger-looking skin & SPF 18 \textbf{sunscreen} protection, broad spectrum UVA/UVB \\ \\
twin bed with \textbf{bookcase} headboard with drawers & Hodedah Twin-Size Captain Bed with 3-Drawers and Headboard in Black & Complete bed plus \textbf{bookcase} headboard to store all those frequently used items. No box spring is necessary, just top with a twin-size mattress and you're ready for bed. Constructed of durable wood and engineered wood. Metal roller glides and built-in safety stop on drawers. The bed can be used with or without the headboard, and drawers can be assembled on either the left or right side. Available in a variety of finishes: Beech, Cherry, Mahogany, Black, Pink, and White. Choose the one that matches your room décor & complete bed plus \textbf{bookcase} headboard twin size no box spring, metal roller glides, wood \\
\bottomrule
\end{tabular}
\caption{Some examples showing how a typical generated summary from ReLSum$_{\text{GRPO}}$ looks like and the corresponding queries that the summary helps with. The examples are taken from our golden test dataset.}
\label{tab:egs}
\end{table*}

\subsection{Examples}

Table~\ref{tab:egs} provides examples of generated summaries from our proposed ReLSum$_{\text{GRPO}}$ framework, showcasing how the approach effectively condenses product descriptions into concise, query-relevant attributes. Each example highlights how the generated summary can help improve ranking when used alongside the product title.

For instance, in the first example, the query ``loreal revitalift sunscreen" matches the product  ``Revitalift Anti-Wrinkle Firming Day Cream by LOreal Paris for Women - 1.7 oz Cream." However, the title alone does not explicitly mention the critical attribute ``sunscreen," which is essential for relevance. The full product description contains this information, but its verbosity makes it impractical for real-time ranking. The generated summary, ``SPF 18 sunscreen protection, broad spectrum UVA/UVB," succinctly captures the key attribute, ensuring the product is correctly ranked for the query.

Similarly, in the second example, the query ``twin bed with bookcase headboard with drawers" partially matches the product title ``Hodedah Twin-Size Captain Bed with 3-Drawers and Headboard in Black." While the title mentions ``headboard" and ``drawers," it does not explicitly highlights the ``bookcase" feature, which is critical for matching the query intent. The generated summary, ``complete bed plus bookcase headboard twin size no box spring, metal roller glides, wood," includes this missing attribute.

The tokens in the generated summaries address false negative problems in ranking by ensuring that critical attributes explicitly mentioned in the query but missing from the product title are included in the ranking context. This reduces the likelihood of relevant products being incorrectly excluded due to incomplete information in the title, thereby improving the accuracy and robustness of the ranking system.

\section{Conclusion and Future Work}
\label{sec:conclude}
In this paper, we proposed a novel reinforcement learning framework, ReLSum, for summarizing product descriptions to optimize search relevance in e-commerce applications. By leveraging relevance scores as rewards, our approach effectively aligns the objectives of summarization and ranking, addressing key limitations of prior methods such as misaligned learning targets and latency constraints. Experimental evaluations demonstrated significant improvements in both offline and online metrics, highlighting the practical benefits of our approach in real-world production systems.

For future work, we aim to explore an iterative training framework where the LLM is fine-tuned based on the proposed approach, followed by fine-tuning the parameters of the downstream relevance model. This iterative process could further enhance the alignment between the generated summaries and ranking objectives. Additionally, we plan to extend the input to the LLM to include other product attributes, such as product images, to enrich the summarization process and improve relevance predictions for visually-driven queries.

\bibliography{aaai2026}

\end{document}